\documentclass[letter]{aa} % for the letters 
\usepackage{graphicx}
\usepackage{txfonts}
\usepackage{lscape}
\usepackage{longtable}
\usepackage{verbatim}
\usepackage[hidelinks]{hyperref}
\usepackage{xcolor}
\usepackage{subcaption}
\hypersetup{colorlinks, linkcolor={blue!70!black}, citecolor={blue!70!black}, urlcolor={blue!70!black}
}
\begin{document}

   \title{Unveiling the past evolution of the progenitor of the Helmi streams}
   
   \author{T. Ruiz-Lara
          \inst{1, 2}
          \and
          A. Helmi \inst{1}
          \and 
          C. Gallart \inst{3, 4}
          \and
          F. Surot \inst{3, 4}
          \and 
          S. Cassisi \inst{5, 6}%\fnmsep\thanks{Just to show the usage of the elements in the author field}
          }

   \titlerunning{Past evoluation of the Helmi stream progenitor}
   \authorrunning{T. Ruiz-Lara et al.}
   
   \institute{Kapteyn Astronomical Institute, University of Groningen,
          Landleven 12, 9747 AD Groningen, The Netherlands\\
   \email{tomasruizlara@gmail.com}
   \and
   Universidad de Granada, Departamento de F\'isica Te\'orica y del Cosmos, Campus Fuente Nueva, Edificio Mecenas, E-18071, Granada, Spain.
   \and
   Instituto de Astrof\'isica de Canarias, E-38200 La Laguna, Tenerife, Spain
   \and 
   Departamento de Astrof\'isica, Universidad de La Laguna, E-38205 La Laguna, Tenerife, Spain
   \and 
   INAF -- Astronomical Observatory of Abruzzo, via M. Maggini, sn, 64100 Teramo, Italy
   \and
   INFN, Sezione di Pisa, Largo Pontecorvo 3, 56127 Pisa, Italy}

   \date{Received --, --; accepted --, --}

% \abstract{}{}{}{}{} 
% 5 {} token are mandatory
 
  \abstract
  % context heading (optional)
  % {} leave it empty if necessary  
   {}
  % aims heading (mandatory)
   {We aim to determine unique features that characterise the past evolution  of the progenitor of the Helmi streams through the analysis of star formation histories (SFHs).}
  % methods heading (mandatory)
   {From the 5D {\it Gaia} EDR3 dataset, we extracted local samples of stars dominated by the Helmi streams, the Galactic (thick and thin) disc, and the local retrograde halo. We did this by identifying regions in a pseudo-Cartesian velocity space (obtained by setting line-of-sight velocities to zero), where stars belonging to these components, as identified in samples with 6D phase-space information, are predominantly found. We made use of an updated absolute colour-magnitude diagram (CMD) fitting methodology to contrast the SFHs of these samples to unveil distinct signatures of the past evolution of a disrupted accreted system: the Helmi streams. To this end, special attention was given to the correct characterisation of {\it Gaia} completeness effects and observational errors on the CMD. We further investigated alternative sample selections to improve the purity of our 5D Helmi stream catalogues.}
  % results heading (mandatory)
   {We find that the progenitor of the Helmi streams experienced an early star formation that was sustained for longer (until 7--9~Gyr ago) than for the Milky Way halo (10--11~Gyr ago). As a consequence, half of its stellar mass was in place $\sim$~0.7~Gyr later. The quenching of star formation in the Helmi stream progenitor   $\sim$~8~Gyr ago suggests it was accreted by the Milky Way around this time, in concert with previous estimates based on the dynamics of the streams.}
  % conclusions heading (optional), leave it empty if necessary 
   {}


   \maketitle
%
%-------------------------------------------------------------------

\section{Introduction}

Galaxies are the outcome of the complex interplay between internal, secular processes and the accretion of external stellar systems and intergalactic gas \citep[e.g.][]{2004ARA&A..42..603K, 2008A&ARv..15..189S, 2012MNRAS.419.3200H}. Galaxy accretion, in particular, is a very defining process. Mergers can both drive morphological and dynamical changes in the host galaxy and trigger the formation of new stars \citep[e.g.][]{1988ApJ...331..699B, 2019A&A...631A..51P}. 

Stellar debris from these merging events, together with heated-up, in situ stars, are deposited in the galactic haloes \citep[][]{1978ApJ...225..357S, 2005ApJ...635..931B,2009ApJ...702.1058Z}, making them crucial for unraveling the ancient history of galaxies, including our own, the Milky Way (MW). In fact, the thorough characterisation of the stellar content of the MW halo has allowed many of our Galaxy's building blocks to be unveiled \citep[e.g.][]{1994Natur.370..194I, 1999Natur.402...53H, 2018Natur.563...85H, 2018MNRAS.478..611B, 2018MNRAS.478.5449M, 2019A&A...631L...9K, 2020ApJ...901...48N, 2021MNRAS.500.1385H}. Thanks to current and ongoing large ground- and space-based surveys \citep[e.g.][]{2021A&A...649A...1G, 2019RAA....19...75L, 2020ApJS..249....3A, 2020AJ....160...82S, 2021MNRAS.506..150B}, we are now in a position to characterise the systems from which our own Galaxy grew \citep[][]{2019NatAs...3..932G, 2019MNRAS.487L..47V, 2021ApJ...908L...8A, 2021arXiv211115423M}.

The identification of stars linked to each of the MW building blocks is commonly done based on the dynamical properties of halo stars, such as integrals of motion or action space \citep[][]{2000MNRAS.319..657H, 2018MNRAS.478.5449M, 2022A&A...665A..57L}. In some cases, chemical information is added \citep[][]{2020ApJ...901...48N, 2022arXiv220102405R}. This approach requires the precise determination in three dimensions of both stellar positions and velocities. Unfortunately, although large samples of stars with known 6D phase-space information have become available with the advent of {\it Gaia} data \citep[][]{2021A&A...649A...1G}, the number of halo stars 
in these samples is still very limited. This, together with completeness and selection effects \citep[][]{2022MNRAS.509.6205E} that especially affect faint magnitudes, considerably limits the characterisation of the hitherto known building blocks of our Galaxy. 

In this work we exploit an alternative approach to isolating samples of stars possibly associated with the Helmi streams \citep[][]{1999Natur.402...53H}
in an attempt to reconstruct the star formation history (SFH) of their progenitor. To this end, we use 5D information, namely their position on the sky, distance, and proper motions from {\it Gaia}. The Helmi streams are debris from a galaxy ($\sim10^8 M_\odot$ in stars) accreted between 5 and 8~Gyr ago \citep[e.g.][]{2007AJ....134.1579K, 2019A&A...625A...5K}. Recent work has reported chemical patterns in the streams' stars that distinguish them from other halo stars \citep[][]{2021MNRAS.500..889A, 2021A&A...651A..57N, 2022arXiv220311808M}, suggesting different formation histories. We find compelling evidence that the progenitor dwarf galaxy continued forming stars for longer than the average halo near the Sun, until it stopped approximately 7 to 9 Gyrs ago, possibly due to its accretion onto the MW.
\section{Data and sample selection}
\label{sec:data}

We considered stars from {\it Gaia} Early\ Data Release 3 \citep[EDR3;][]{2021A&A...649A...1G} with \texttt{parallax\_over\_error}~$>5$ and good \texttt{phot\_bp\_rp\_excess\_factor} \citep[see][]{2022arXiv220102405R} and which are located within 2.5 kpc of the Sun, as determined by inverting their parallax after applying a global zero-point offset \citep{2021A&A...649A...4L}.  For all stars in this `local' sample, we computed their absolute colour ($G_{\rm BP}-G_{\rm RP}$) and {\it G} magnitude ($M_G$) using their parallax, and corrected for extinction (E(BP-RP) and A$_G$), from \citet{2018JOSS....3..695M} and \citet{2018JOSS....3..695M2019ApJ...887...93G}, with the recipes presented in \citet[][]{2018A&A...616A..10G}. 

To guide our selection of Helmi stream stars in 5D, we first identified a sample of halo stars with 6D information. This was obtained for the local sample by complementing the radial velocities from {\it Gaia} EDR3 with $v_{los}$ measurements from GALAH Data Release (DR) 3 \citep{2021MNRAS.506..150B}, APOGEE DR16 \citep{2020ApJS..249....3A}, RAVE DR6 \citep{2020AJ....160...82S, 2020AJ....160...83S}, and LAMOST DR6 \citep{2019RAA....19...75L, 2020ApJS..251...27W}. Velocity systematic shifts between surveys, although small \citep[see][]{2022A&A...659A..95T}, were considered. We selected halo stars by requiring that  $|{\bf V}-{\bf V}_{LSR}|$~>~210~km/s, where ${\bf V}$ is the total velocity vector corrected for the solar motion and local standard of rest velocity (V$_{LSR}$ = 232~km/s from \citealt{2017MNRAS.465...76M}). 

We identified 646 candidate members of the Helmi streams in this 6D halo sample using the criteria from \citet{2019A&A...625A...5K}, \citet[][]{2022A&A...659A..61D}, and \citet[][]{2022A&A...665A..57L}.  Ideally, one would compute the SFH of the progenitor of the Helmi streams  from this subset. Unfortunately, the limited number of stars available and especially the complex selection functions (from {\it Gaia} as well as the various spectroscopic surveys used) hinder this approach. 

This is why we defined a 5D halo sample. It was extracted from the local sample as follows. For stars near the Galactic plane ($|b|<20^{\circ}$), we considered candidate halo stars as those with a tangential velocity v$_t$~=~4.74/$\texttt{parallax} \times (\mu_{\alpha*}^2+\mu_{\delta}^2)^{1/2} > 230$~km/s. For stars with  $|b|>20^{\circ}$, we assigned the v$_{los}$ from the star in the 6D {\it Gaia} ({local}) sample that is closest in the space of $(\alpha, \delta, \texttt{parallax})$, thus obtaining a fictitious ${\bf V'}$. Then, we applied the same criterion as in 6D, namely $|{\bf V'}-{\bf V}_{LSR}|$~>~210~km/s. In both cases, we kept only stars with $M_G$ < 5.

\begin{figure}
\centering
\includegraphics[width=0.45\textwidth]{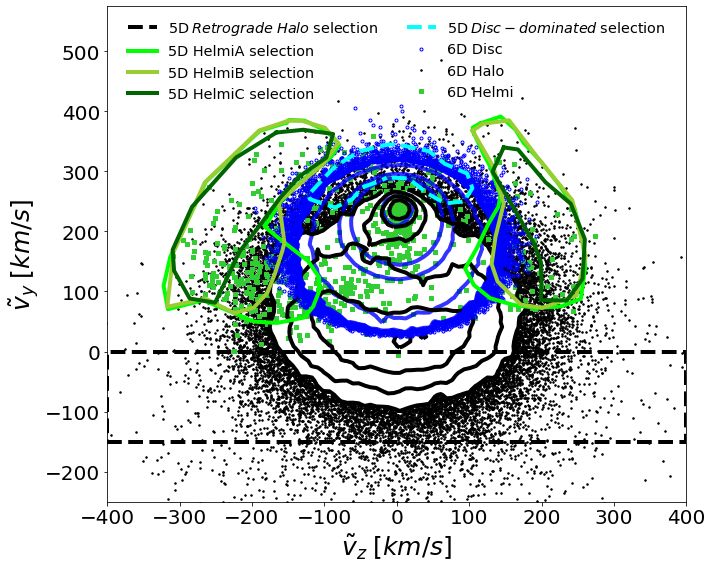}
\caption{Selection of samples in the pseudo-Cartesian velocity plane, $\tilde{v}_y$ vs $\tilde{v}_z$. Black contours and points show the location of nearby halo stars identified in 6D (see text for details), and blue represents those of the disc(s). Helmi stream stars selected using 6D information are shown as green points (see text for details). The different 5D Helmi stream selections are shown as solid polygons in hues of green, whereas the {`local retrograde halo'} and the {`disc(s)-dominated'} subsets are given by the dashed black and cyan polygons.
\label{fig:sample}}
\end{figure}

To select tentative Helmi member stars from this 5D halo sample, we used their pseudo-Cartesian velocities. These were obtained by computing the velocities in Cartesian coordinates assuming $v_{los} = 0$ \citep[see Eq.~6 of ][for the exact expressions]{2021A&A...649A.136K}.  The pseudo-Cartesian velocities offer an improvement over a selection based on proper motions only (or projected velocities) because they take their dependence on sky location into account. Figure~\ref{fig:sample} displays the distribution of the Helmi stream stars from the 6D subsets in this pseudo-Cartesian velocity space. Although the distinctive clustering in $v_y$ versus~$v_z$ that led to the discovery of the Helmi streams is less prominent in this pseudo-space, some differentiation is still possible as there are clear regions in  $\tilde{v}_{y}$ versus~$\tilde{v}_{z}$ where the stream stars are more dominant.

We thus proceeded to define three different sub-samples from the 5D halo: HelmiA ($\sim$48000), HelmiB ($\sim$23000), and HelmiC ($\sim$7000 stars, see Fig.~\ref{fig:sample}); we kept only stars with A$_G<0.5$ as they will be used for the computation of SFHs \citep[][]{2020NatAs...4..965R}. HelmiC is the strictest subset and, thus, has the least amount of contamination. 

We estimated the completeness and purity of these subsets as follows. We compared the number of members of the Helmi streams (identified in 6D) in the subsets HelmiA, HelmiB, and HelmiC, $N^{\rm mem}_{\rm A/B/C}$, to their total number in our 5D halo sample, $N^{\rm mem}_{5D}$ (as this is the maximum number of Helmi stream members that can enter our subsets). The completeness is thus $N^{\rm mem}_{\rm A/B/C}/N^{\rm mem}_{5D}$, yielding 41\%, 28\%, and 12\%, respectively\footnote{This would be the completeness if the {\it Gaia} RV (Radial Velocity) sample were complete; however, because of selection effects, these values should be interpreted as lower limits.}. To estimate the purity of the subsets, we compared the relative number of true members, $N^{\rm mem}_{\rm A/B/C}$, to the number of stars from the 6D ({\it Gaia}) halo  sample in each subset, $N^{\rm 6D}_{\rm A/B/C}$. We find that HelmiA, HelmiB, and HelmiC have a purity of 3.2\%, 8.5\%, and 14\%, respectively, confirming the lower contamination of HelmiC. From the ({\it Gaia}) 6D halo sample we can also determine which other structures fall in the region of the HelmiC subset. We find that the main source of contamination are stars from Gaia-Enceladus  and from the hot thick disc. 

We constructed an additional subset based on HelmiA as this set has the highest completeness. This new subset, HelmiA$^*$, was obtained as follows. To each star in HelmiA, we assigned the v$_{los}$ of the closest star in the ($\mu_l$, $\mu_b$, $l$, $b$, distance) space from the 6D ({\it Gaia}) halo  sample. We then computed their integrals of motion  ($L_{z}$, $L_{\perp}$, and energy) and only kept stars with values compatible with belonging to the Helmi streams \citep[following][]{2022A&A...659A..61D, 2022A&A...665A..57L}. By definition, the purity of this sample is 100\% (as all the stars in HelmiA$^*$ will have both proper motions and line-of-sight velocities that make them true members of the Helmi streams\footnote{Note, however, that the sample may still include stars from a background population that has similar integrals of motion.}). Its completeness is 23.9\%, which is thus higher than that of HelmiC. HelmiA$^*$ contains 1444 stars, 801 after quality cuts are applied. 

To aid our understanding of the SFHs that we will infer in the next section, we selected a sample representative of the local retrograde halo by considering stars with negative $\tilde{v}_y$ (`local retrograde halo'; $\sim$96000 stars). We also defined a `(thin + thick) disc-dominated' sample ($\sim$12000 stars), where disc stars seem to dominate in pseudo-Cartesian velocity space as inferred from comparison to a 6D sample (defined by stars with $|{\bf V}-{\bf V}_{LSR}| <210$~km/s). We expect a comparison of the SFHs derived for the various sets to reveal intrinsic differences in the formation histories of the MW local retrograde halo, the disc, and the progenitor of the Helmi streams.

\section{Methodology}
\label{sec:methods}

Fitting an absolute colour-magnitude diagram\footnote{In what follows, although we speak of CMD only, it will be defined in the absolute plane, not the apparent one.} (CMD) has proven to be an efficient way of retrieving SFHs of stellar systems, including our own Galaxy \citep[e.g.][]{1999AJ....118.2245G, 2019NatAs...3..932G, tolstoy2009, cignonitosi2010, 2020NatAs...4..965R}. In this work we used an updated CMD fitting methodology tailored for {\it Gaia} data that we call `CMDft.Gaia'\footnote{CMDft.Gaia is a suite of procedures that includes: (i) the computation of synthetic CMDs; (ii) the simulation of observational errors and completeness \citep[DisPar-Gaia; see also][]{2021MNRAS.501.3962R}; and (iii) the proper derivation of the SFH (dirSFH).}. All the details are reported in Gallart et al. (in prep.). Here, we provide a brief description.  

\begin{figure*}[h]
\centering
\includegraphics[width=0.9\textwidth]{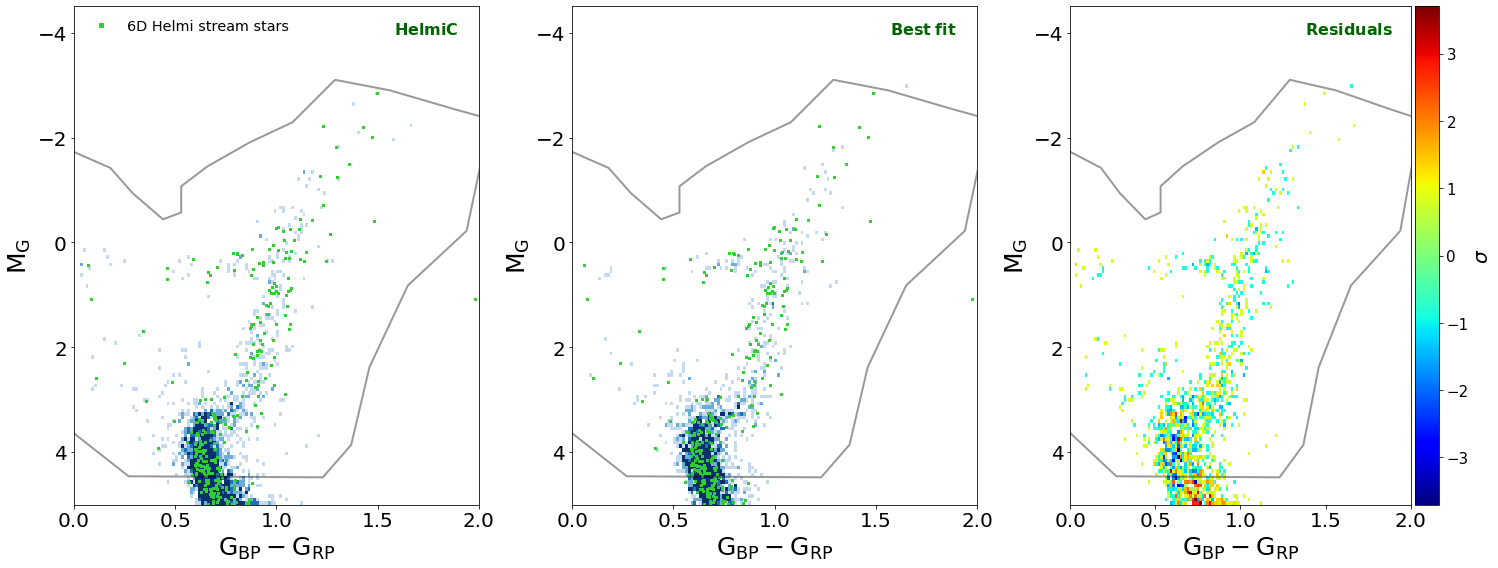}
\caption{Results of the absolute CMD fitting procedure (CMDft.Gaia) for the tentative members of the Helmi streams selected in 5D and part of the HelmiC subset. The left and middle panels show, respectively, the distribution of observed stars in the subset and the best fit. The green points correspond to the Helmi stream stars selected in 6D. The right-hand panel displays the residuals of the CMD fitting (in $\sigma$ units assuming Poissonian errors).
\label{fig:CMD_fit}}
\end{figure*}

\subsection{Synthetic CMD: Completeness and error simulation}
\label{dispar}

We compared the various 5D samples selected in Sect.~\ref{sec:data} with a synthetic CMD that contains 40 million stars (with $-3 < M_G<$~5) with a flat distribution of age and metallicity (Z) ranging from 0.02 to 13.5 Gyr and 0.0001 to 0.032, respectively. We computed this synthetic CMD using the updated BaSTI stellar evolutionary models \citep[][]{2018ApJ...856..125H} in the $\alpha$-enhanced version \citep[][]{2021ApJ...908..102P}, with a Reimers mass loss parameter ($\eta$) of 0.3, assuming a Kroupa initial mass fraction \citep[][]{kroupa2001}, a fraction of unresolved binaries ($\beta$) of 30\%, and a minimum mass ratio for binaries ($q$) of 0.1. 

The full {\it Gaia} 5D dataset has unprecedented photometric precision and is basically complete in the volume and absolute colour-magnitude range covered in this work. In particular, within a parallax (or distance cut) of 1/2.5~mas, the set reaches a completeness of $> 99$\% for absolute magnitudes of $-3 < M_G<5$, that is, below the oldest main sequence turnoff \cite[see][]{2019A&A...624L...1M,2022MNRAS.509.6205E}. However, the subsets identified in the previous section were subject to several quality cuts, and this may affect the distribution of stars in the CMD, potentially biasing our inference of SFHs.  

To account for the associated selection effects, we constructed `parent' sub-samples intended to be analogues of the HelmiA, A$^*$, B, and C, local retrograde halo, and disc-dominated 5D subsets identified in Sect.~\ref{sec:data}. They were extracted from the 5D {\it Gaia} sample in the 2.5~kpc volume using the same approach as before (see Fig.~\ref{fig:sample}) but without imposing any quality cuts. As in Sect.~\ref{sec:data}, we computed a tangential velocity, $v_t$, and a fictitious ${\bf V'}$ by inverting the parallax to infer a distance. We note that computing a distance without an error cut could lead to incorrect distances, but this is exactly one of the effects we wanted to simulate.

We simulated the effect of the quality cuts as well as photometric errors as follows (DisPar-Gaia; Gallart et al. in prep.). To each synthetic star we assigned $l$-$b$-\texttt{parallax} based on the global distribution of the
corresponding parent sub-sample (be it the counterpart of the
HelmiA, HelmiA$^*$, HelmiB, HelmiC,  local retrograde halo, or  disc-dominated subsets). This allowed us to shift the synthetic CMD to an apparent CMD and compute the extinction for each star
\citep[][]{2019ApJ...887...93G}. In a second step, we assigned values of the \texttt{phot\_bp\_rp\_excess\_factor},
\texttt{parallax\_over\_error}, and photometric errors (in the three
{\it Gaia} bands) to each
synthetic star from an observed star with a similar absolute colour and magnitude (within 0.02
magnitudes). Sometimes such a
counterpart does not exist; for example, if the observed population is
predominantly old, there will not be young bright main sequence stars,
which, however, are generated in the synthetic CMD. In this case, we
simulated these properties by fitting how the quality parameters vary
as a function of colour and apparent magnitudes for the parent sub-sample.  Finally, new, error-convolved absolute colours and magnitudes were computed for each synthetic star considering the
attributed photometric and parallax errors and extinction
values \citep[][]{2019ApJ...887...93G}. This allowed
us to mimic the effect of the observational errors that blur features in
the CMD.

After simulating the effect of errors, we considered the issue of completeness in two
separate steps. First, although nearly negligible for the
volume and the colour-magnitude ranges considered here, we
used the {\it Gaia} selection function
\citep[][]{2022MNRAS.509.6205E} to evaluate whether a given synthetic
star (based on its assigned $l$, $b$, and simulated apparent $G$ magnitude) would
be included in the 5D {\it Gaia} EDR3 catalogue. Secondly, we
applied the same quality cuts described in Sect.~\ref{sec:data}
(on \texttt{parallax\_over\_error},
\texttt{phot\_bp\_rp\_excess\_factor}, and A$_G$). The outcome of this
procedure is an error-convolved synthetic CMD that is affected by {\it Gaia} observational and selection effects in a similar way as our observed
sub-samples.

\begin{figure*}%[h]
\centering
\includegraphics[width=0.95\textwidth]{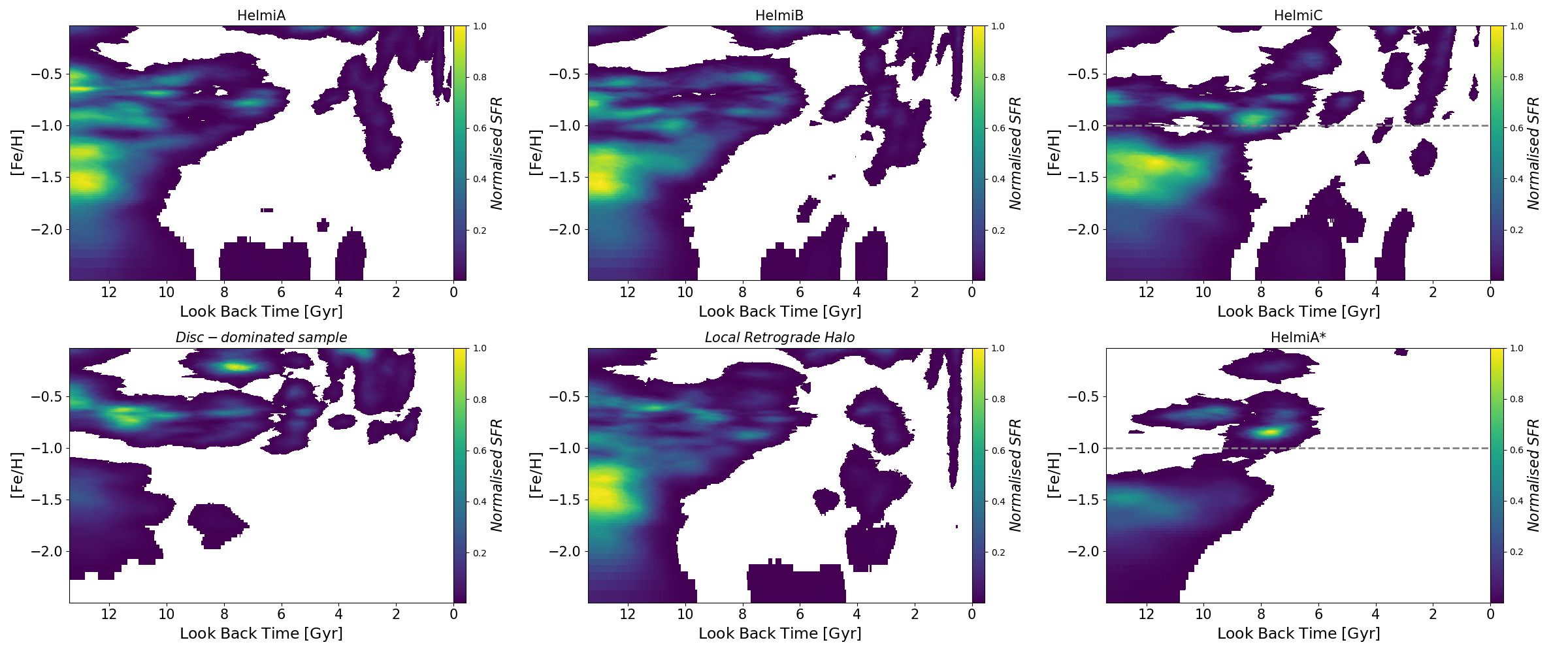}
\caption{Normalised star formation rate in the age-[Fe/H] plane derived from {CMDft.Gaia} for the various samples analysed in this work, namely HelmiA, HelmiB, and HelmiC  (top row) and the {`disc-dominated' subset,}  {the `local retrograde halo'} subset, and HelmiA$^*$ (bottom row). In the right-most panels, the horizontal dashed line tentatively separates the Helmi stream progenitor population from  contamination by the hot thick disc (at [Fe/H]~=~$-1$).
\label{fig:2D_SFH}}
\end{figure*}

\subsection{Computing star formation histories}
\label{sec:dirSFH}

The derivation of the SFHs for our various sub-samples is done using dirSFH (Gallart et al. in prep.), which is an improvement and extension of the well-known tools IACpop \citep{apariciohidalgo2009} and TheStorm \citep[][]{bernard2018MNRAS}. In short, dirSFH defines a series of simple stellar populations from the error-convolved synthetic CMD using a {\it dir}ichlet tessellation
\citep[][]{10.1093/comjnl/21.2.168} from a grid of seed points within the available range of ages and metallicities. The code then finds the combination of simple stellar populations that best fits the observed CMD based on the Skellam probability distribution of the difference between two statistically independent distributions (observed and simulated CMDs). It includes two different weighting strategies as a function of colour and magnitude, namely `uniform' or `weighted' (as the logarithm of the inverse of the variance of the ages across the synthetic CMD). As this weighting scheme already gives which parts of the CMD provide more information for the recovery of the SFH, a single region, or `bundle', encompassing the whole CMD was used (see Fig.~\ref{fig:CMD_fit}), in contrast to, for example, \citet[][]{monelli2010}. The final SFH was derived from the weighted average of the 100 individual solutions that were obtained by slightly modifying each time the grid of seed points and, thus, the tessellation in age and metallicity. The uncertainties were derived directly from the variance of the combination. Extensive testing using different synthetic CMDs (stellar models, unresolved binary recipes, etc.) and dirSFH internal parameters (age-metallicity grids, weighting strategies, etc.) reveal that the solutions are robust (see Appendix~\ref{app:tests}). 

Figure~\ref{fig:CMD_fit} shows the CMD fitting approach -- CMDft.Gaia --  applied to the HelmiC sub-sample. The residuals of the fit (right-most panel) are small and homogeneous across the whole CMD, indicative of a very good fit. We note how well the distribution of stars in the CMD of the 5D-defined HelmiC sub-sample (and its best fit) compares to the CMD of 6D Helmi stream stars.

\section{Results and discussion}
\label{sec:results}

Figure~\ref{fig:2D_SFH} shows the normalised star formation rate as a function of age and [Fe/H] for the various sub-samples: HelmiA, HelmiB, HelmiC, the disc-dominated subset, and the local retrograde halo together with the HelmiA$^*$. Comparing the top panels for the Helmi streams (in decreasing order of
contamination), we see two main trends: (i) stars older than 8~Gyr
with [Fe/H]~$\in (-1.0,-0.4)$ tend to be less dominant, and almost
absent in HelmiC; and (ii) the metal-poor population ([Fe/H]$\sim-1.5$) gradually extends to younger ages (from up to 11 Gyr in
HelmiA to nearly 9~Gyr in HelmiC).  Trend (i) is a direct
consequence of the HelmiB and HelmiC sub-samples presenting a higher purity and a lower amount of contamination from (thick) disc stars (see
Fig.~\ref{fig:sample}), as a comparison to the bottom-left panel reveals. Point (ii) suggests that the Helmi
stream progenitor experienced early star formation that extended to younger ages than the populations contaminating these samples.  

\begin{figure}
\centering
\includegraphics[width=0.4\textwidth]{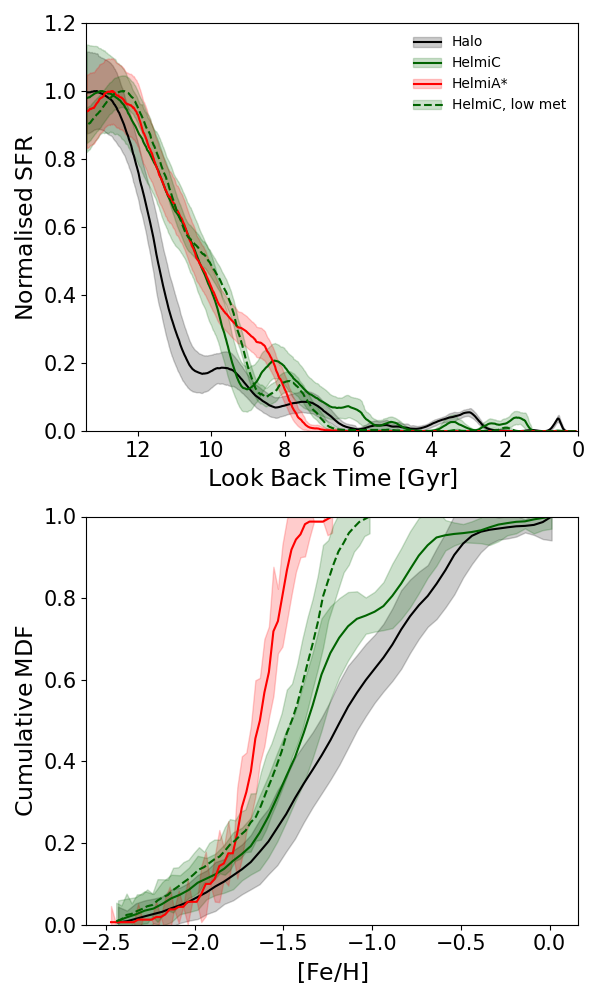}
\caption{Comparison of the SFHs of the HelmiC (in green) and  HelmiA$^*$ (in red) subsets with that of the local retrograde halo sample (black). {\it Top:} Normalised star formation rate as a function of age. {\it Bottom:} Cumulative metallicity distribution function (MDF). The SFHs in the upper panel have been normalised to their peak value. Note the slightly larger uncertainties for the HelmiA$^*$ subset, which are due to its smaller size. For  HelmiA$^*$ (and the HelmiC-low met subset in the dashed green line), we  restrict the comparison to the population with [Fe/H]~$\le -1.0$ (to remove the contamination by the disc).}
\label{fig:Helmi_halo_comp}
\end{figure}

The bottom-middle panel of Figure~\ref{fig:2D_SFH} for the local retrograde halo shows this is dominated by old and metal-poor stars, in agreement with current knowledge \citep{2020ARA&A..58..205H}.  The extension that is seen to higher metallicities and young ages is possibly a consequence of disc star contamination, as a comparison to the bottom-left panel shows. We verified that these results are independent of the exact selection of the halo sample.

The derived SFH from the HelmiA$^*$ subset (bottom-right panel of Fig.~\ref{fig:2D_SFH}) shows a `cleaned-up' version of what we find for HelmiC, as expected from its higher purity and completeness. While the HelmiC sample is  mainly contaminated by the hot thick disc and {\it Gaia}-Enceladus (as discussed in Sect.~\ref{sec:data} and confirmed from a comparison of the various panels in Fig.~\ref{fig:2D_SFH}), HelmiA$^*$ appears to be mainly contaminated by hot thick disc stars, as $\sim$25\% of the stars have [Fe/H]~$> -1.0$  in the solution shown in Fig.~\ref{fig:2D_SFH}. As a consequence,  genuine features of the past evolution of the Helmi stream progenitor are more easily discernable in this sample, and we can confirm that it formed stars for longer than the local halo.

Figure~\ref{fig:Helmi_halo_comp} compares more directly the SFH of the HelmiC and HelmiA$^*$ subsets to that representing the local retrograde halo. In the top panel we can clearly see that HelmiC and HelmiA$^*$ display a SFH that is more extended in time, with star formation being quenched\footnote{Normalised star formation rate below 0.2, 20$\%$.} $\sim$~7 to 9~Gyrs ago, 2 to 3~Gyr later than the halo sample.  Our findings also suggest that the progenitor of the Helmi streams formed 50$\%$ of its stellar mass $\sim$~0.7~Gyr later ($\sim$~11.5~Gyr ago) than the  local retrograde halo sample, and up to $\sim$~1.0~Gyr later if we consider the HelmiA$^*$ subset. The bottom panel of Fig.~\ref{fig:Helmi_halo_comp} shows that the HelmiC and HelmiA* subsets present lower median metallicities than the comparison halo sample, in agreement with previous work. Some hints of a star formation burst at $\sim$8~Gyr ago can be seen in the HelmiC sample (probably triggered during the merging process of the Helmi stream progenitor with the MW), although it is minimised in the case of the HelmiA* sample.

\citet[][]{2019A&A...625A...5K}, comparing isochrones to Helmi stream
stars in the CMD, suggested a spread in age from at least $\sim$~11 to 13~Gyr,
and their N-body analysis favoured an accretion time between 5 and
8~Gyr ago \citep[see also][]{2007AJ....134.1579K}. In this sense, the
clear quenching of star formation detected in the HelmiC and HelmiA$^*$ subsets
between $\sim$~7 to 9~Gyr ago might be interpreted as a sign of
accretion happening at that point. This is in remarkable agreement with the quenching time estimated by \citet[][]{2022arXiv220409057N}, who dated the accretion to z$\sim$1 ($\sim$7~Gyr ago) based on the $\alpha$ and Fe abundances of the stars. The low and constant
value of [$\alpha$/Fe] of Helmi stream stars reported in
\citet[][see also \citealt{2010ApJ...711..573R}; \citealt{2021ApJ...912...52G}; \citealt{2021ApJ...913L..28L}; or \citealt{2022arXiv220404233H}]{2022arXiv220311808M} has been interpreted by these authors as an indication that its progenitor should
have had a quiescent and extended SFH, and may have
experienced a small burst at late times, results that are consistent with our findings.

\section{Conclusions}

In this Letter we have unveiled some  characteristics of the past evolution of the progenitor of the Helmi streams using a sample of nearby tentative members extracted from the 5D {\it Gaia} EDR3 dataset. We have found that the progenitor of the streams displayed a more extended early star formation than a comparison sample representative of the local, retrograde, MW halo, as well as an average lower metallicity. In addition, we have clearly detected the quenching of its star formation $\sim$~7 to 9~Gyr ago, which may have coincided with its accretion time.  The availability of line-of-sight velocities for larger samples of stars expected in future {\it Gaia} data releases and upcoming spectroscopic surveys will soon enable the determination of SFHs for much purer samples of stars possibly associated with the MW building blocks. This will provide an unprecedented, time-resolved picture of the accretion history of our Galaxy.

\begin{acknowledgements}
We are grateful to the anonymous referee for providing key comments that greatly improved the quality of this work. We  gratefully  acknowledge  financial  support from a Spinoza prize. TRL acknowledges support from project PID2020-114414GB-100 and Juan de la Cierva fellowship (IJC2020-043742-I), financed by MCIN/AEI/10.13039/501100011033. SC acknowledges support from the Fundaci\'on Jes\'us Serra and the Instituto de Astrof\'isica de Canarias under the Visiting Researcher Programme 2020-2022 agreed between both institutions.
This work has made use of data from the European Space Agency (ESA) mission {\it Gaia} (\url{https://www.cosmos.esa.int/gaia}), processed by the {\it Gaia} Data Processing and Analysis Consortium (DPAC, \url{https://www.cosmos.esa.int/web/gaia/dpac/consortium}). Funding for the DPAC has been provided by national institutions, in particular the institutions participating in the {\it Gaia} Multilateral Agreement.
This research makes use of {\tt python}; {\tt vaex} \citep[][]{2018A&A...618A..13B}, a {\tt Python} library for visualization and exploration of big tabular data sets; {\tt matplotlib} \citep[][]{hunter2007}, a suite of open-source python modules that provide a framework for creating scientific plots; {\tt astropy} \citep[][]{2018AJ....156..123A}, a community-developed core {\tt Python} package for Astronomy; {\tt numpy}; and {\tt jupyter notebooks}.
This paper has made use of data from GALAH, APOGEE, RAVE, and LAMOST spectroscopic surveys.
\end{acknowledgements}

\bibliographystyle{aa} % style aa.bst
\bibliography{references.bib}

\begin{thebibliography}{68}
\expandafter\ifx\csname natexlab\endcsname\relax\def\natexlab#1{#1}\fi

\bibitem[{{Aguado} {et~al.}(2021{\natexlab{a}}){Aguado}, {Belokurov}, {Myeong},
  {Evans}, {Kobayashi}, {Sbordone}, {Chanam{\'e}}, {Navarrete}, \&
  {Koposov}}]{2021ApJ...908L...8A}
{Aguado}, D.~S., {Belokurov}, V., {Myeong}, G.~C., {et~al.} 2021{\natexlab{a}},
  \apjl, 908, L8

\bibitem[{{Aguado} {et~al.}(2021{\natexlab{b}}){Aguado}, {Myeong}, {Belokurov},
  {Evans}, {Koposov}, {Allende Prieto}, {Lanfranchi}, {Matteucci}, {Shetrone},
  {Sbordone}, {Navarrete}, {Gonz{\'a}lez Hern{\'a}ndez}, {Chanam{\'e}},
  {Peralta de Arriba}, \& {Yuan}}]{2021MNRAS.500..889A}
{Aguado}, D.~S., {Myeong}, G.~C., {Belokurov}, V., {et~al.} 2021{\natexlab{b}},
  \mnras, 500, 889

\bibitem[{{Ahumada} {et~al.}(2020){Ahumada}, {Prieto}, {Almeida}, {Anders},
  {Anderson}, {Andrews}, {Anguiano}, {Arcodia}, {Armengaud}, {Aubert}, {Avila},
  {Avila-Reese}, {Badenes}, {Balland}, {Barger}, {Barrera-Ballesteros}, {Basu},
  {Bautista}, {Beaton}, {Beers}, {Benavides}, {Bender}, {Bernardi}, {Bershady},
  {Beutler}, {Bidin}, {Bird}, {Bizyaev}, {Blanc}, {Blanton}, {Boquien},
  {Borissova}, {Bovy}, {Brandt}, {Brinkmann}, {Brownstein}, {Bundy}, {Bureau},
  {Burgasser}, {Burtin}, {Cano-D{\'\i}az}, {Capasso}, {Cappellari}, {Carrera},
  {Chabanier}, {Chaplin}, {Chapman}, {Cherinka}, {Chiappini}, {Doohyun Choi},
  {Chojnowski}, {Chung}, {Clerc}, {Coffey}, {Comerford}, {Comparat}, {da
  Costa}, {Cousinou}, {Covey}, {Crane}, {Cunha}, {Ilha}, {Dai}, {Damsted},
  {Darling}, {Davidson}, {Davies}, {Dawson}, {De}, {de la Macorra}, {De Lee},
  {Queiroz}, {Deconto Machado}, {de la Torre}, {Dell'Agli}, {du Mas des
  Bourboux}, {Diamond-Stanic}, {Dillon}, {Donor}, {Drory}, {Duckworth},
  {Dwelly}, {Ebelke}, {Eftekharzadeh}, {Davis Eigenbrot}, {Elsworth},
  {Eracleous}, {Erfanianfar}, {Escoffier}, {Fan}, {Farr},
  {Fern{\'a}ndez-Trincado}, {Feuillet}, {Finoguenov}, {Fofie},
  {Fraser-McKelvie}, {Frinchaboy}, {Fromenteau}, {Fu}, {Galbany}, {Garcia},
  {Garc{\'\i}a-Hern{\'a}ndez}, {Oehmichen}, {Ge}, {Maia}, {Geisler}, {Gelfand},
  {Goddy}, {Gonzalez-Perez}, {Grabowski}, {Green}, {Grier}, {Guo}, {Guy},
  {Harding}, {Hasselquist}, {Hawken}, {Hayes}, {Hearty}, {Hekker}, {Hogg},
  {Holtzman}, {Horta}, {Hou}, {Hsieh}, {Huber}, {Hunt}, {Chitham}, {Imig},
  {Jaber}, {Angel}, {Johnson}, {Jones}, {J{\"o}nsson}, {Jullo}, {Kim},
  {Kinemuchi}, {Kirkpatrick}, {Kite}, {Klaene}, {Kneib}, {Kollmeier}, {Kong},
  {Kounkel}, {Krishnarao}, {Lacerna}, {Lan}, {Lane}, {Law}, {Le Goff}, {Leung},
  {Lewis}, {Li}, {Lian}, {Lin}, {Long}, {Longa-Pe{\~n}a}, {Lundgren}, {Lyke},
  {Ted Mackereth}, {MacLeod}, {Majewski}, {Manchado}, {Maraston}, {Martini},
  {Masseron}, {Masters}, {Mathur}, {McDermid}, {Merloni}, {Merrifield},
  {M{\'e}sz{\'a}ros}, {Miglio}, {Minniti}, {Minsley}, {Miyaji}, {Mohammad},
  {Mosser}, {Mueller}, {Muna}, {Mu{\~n}oz-Guti{\'e}rrez}, {Myers}, {Nadathur},
  {Nair}, {Nandra}, {do Nascimento}, {Nevin}, {Newman}, {Nidever}, {Nitschelm},
  {Noterdaeme}, {O'Connell}, {Olmstead}, {Oravetz}, {Oravetz}, {Osorio},
  {Pace}, {Padilla}, {Palanque-Delabrouille}, {Palicio}, {Pan}, {Pan},
  {Parker}, {Paviot}, {Peirani}, {Ram{\'r}ez}, {Penny}, {Percival},
  {Perez-Fournon}, {P{\'e}rez-R{\`a}fols}, {Petitjean}, {Pieri},
  {Pinsonneault}, {Poovelil}, {Povick}, {Prakash}, {Price-Whelan}, {Raddick},
  {Raichoor}, {Ray}, {Rembold}, {Rezaie}, {Riffel}, {Riffel}, {Rix}, {Robin},
  {Roman-Lopes}, {Rom{\'a}n-Z{\'u}{\~n}iga}, {Rose}, {Ross}, {Rossi},
  {Rowlands}, {Rubin}, {Salvato}, {S{\'a}nchez}, {S{\'a}nchez-Menguiano},
  {S{\'a}nchez-Gallego}, {Sayres}, {Schaefer}, {Schiavon}, {Schimoia},
  {Schlafly}, {Schlegel}, {Schneider}, {Schultheis}, {Schwope}, {Seo},
  {Serenelli}, {Shafieloo}, {Shamsi}, {Shao}, {Shen}, {Shetrone}, {Shirley},
  {Aguirre}, {Simon}, {Skrutskie}, {Slosar}, {Smethurst}, {Sobeck}, {Sodi},
  {Souto}, {Stark}, {Stassun}, {Steinmetz}, {Stello}, {Stermer},
  {Storchi-Bergmann}, {Streblyanska}, {Stringfellow}, {Stutz}, {Su{\'a}rez},
  {Sun}, {Taghizadeh-Popp}, {Talbot}, {Tayar}, {Thakar}, {Theriault}, {Thomas},
  {Thomas}, {Tinker}, {Tojeiro}, {Toledo}, {Tremonti}, {Troup}, {Tuttle},
  {Unda-Sanzana}, {Valentini}, {Vargas-Gonz{\'a}lez}, {Vargas-Maga{\~n}a},
  {V{\'a}zquez-Mata}, {Vivek}, {Wake}, {Wang}, {Weaver}, {Weijmans}, {Wild},
  {Wilson}, {Wilson}, {Wolthuis}, {Wood-Vasey}, {Yan}, {Yang}, {Y{\`e}che},
  {Zamora}, {Zarrouk}, {Zasowski}, {Zhang}, {Zhao}, {Zhao}, {Zheng}, {Zheng},
  {Zhu}, \& {Zou}}]{2020ApJS..249....3A}
{Ahumada}, R., {Prieto}, C.~A., {Almeida}, A., {et~al.} 2020, \apjs, 249, 3

\bibitem[{{Aparicio} \& {Hidalgo}(2009)}]{apariciohidalgo2009}
{Aparicio}, A. \& {Hidalgo}, S.~L. 2009, \aj, 138, 558

\bibitem[{{Astropy Collaboration} {et~al.}(2018){Astropy Collaboration},
  {Price-Whelan}, {Sip{\H{o}}cz}, {G{\"u}nther}, {Lim}, {Crawford}, {Conseil},
  {Shupe}, {Craig}, {Dencheva}, {Ginsburg}, {Vand erPlas}, {Bradley},
  {P{\'e}rez-Su{\'a}rez}, {de Val-Borro}, {Aldcroft}, {Cruz}, {Robitaille},
  {Tollerud}, {Ardelean}, {Babej}, {Bach}, {Bachetti}, {Bakanov}, {Bamford},
  {Barentsen}, {Barmby}, {Baumbach}, {Berry}, {Biscani}, {Boquien}, {Bostroem},
  {Bouma}, {Brammer}, {Bray}, {Breytenbach}, {Buddelmeijer}, {Burke},
  {Calderone}, {Cano Rodr{\'\i}guez}, {Cara}, {Cardoso}, {Cheedella}, {Copin},
  {Corrales}, {Crichton}, {D'Avella}, {Deil}, {Depagne}, {Dietrich}, {Donath},
  {Droettboom}, {Earl}, {Erben}, {Fabbro}, {Ferreira}, {Finethy}, {Fox},
  {Garrison}, {Gibbons}, {Goldstein}, {Gommers}, {Greco}, {Greenfield},
  {Groener}, {Grollier}, {Hagen}, {Hirst}, {Homeier}, {Horton}, {Hosseinzadeh},
  {Hu}, {Hunkeler}, {Ivezi{\'c}}, {Jain}, {Jenness}, {Kanarek}, {Kendrew},
  {Kern}, {Kerzendorf}, {Khvalko}, {King}, {Kirkby}, {Kulkarni}, {Kumar},
  {Lee}, {Lenz}, {Littlefair}, {Ma}, {Macleod}, {Mastropietro}, {McCully},
  {Montagnac}, {Morris}, {Mueller}, {Mumford}, {Muna}, {Murphy}, {Nelson},
  {Nguyen}, {Ninan}, {N{\"o}the}, {Ogaz}, {Oh}, {Parejko}, {Parley}, {Pascual},
  {Patil}, {Patil}, {Plunkett}, {Prochaska}, {Rastogi}, {Reddy Janga},
  {Sabater}, {Sakurikar}, {Seifert}, {Sherbert}, {Sherwood-Taylor}, {Shih},
  {Sick}, {Silbiger}, {Singanamalla}, {Singer}, {Sladen}, {Sooley},
  {Sornarajah}, {Streicher}, {Teuben}, {Thomas}, {Tremblay}, {Turner},
  {Terr{\'o}n}, {van Kerkwijk}, {de la Vega}, {Watkins}, {Weaver}, {Whitmore},
  {Woillez}, {Zabalza}, \& {Astropy Contributors}}]{2018AJ....156..123A}
{Astropy Collaboration}, {Price-Whelan}, A.~M., {Sip{\H{o}}cz}, B.~M., {et~al.}
  2018, \aj, 156, 123

\bibitem[{{Barnes}(1988)}]{1988ApJ...331..699B}
{Barnes}, J.~E. 1988, \apj, 331, 699

\bibitem[{{Belokurov} {et~al.}(2018){Belokurov}, {Erkal}, {Evans}, {Koposov},
  \& {Deason}}]{2018MNRAS.478..611B}
{Belokurov}, V., {Erkal}, D., {Evans}, N.~W., {Koposov}, S.~E., \& {Deason},
  A.~J. 2018, \mnras, 478, 611

\bibitem[{{Belokurov} {et~al.}(2020){Belokurov}, {Penoyre}, {Oh}, {Iorio},
  {Hodgkin}, {Evans}, {Everall}, {Koposov}, {Tout}, {Izzard}, {Clarke}, \&
  {Brown}}]{2020MNRAS.496.1922B}
{Belokurov}, V., {Penoyre}, Z., {Oh}, S., {et~al.} 2020, \mnras, 496, 1922

\bibitem[{{Bernard} {et~al.}(2018){Bernard}, {Schultheis}, {Di Matteo}, {Hill},
  {Haywood}, \& {Calamida}}]{bernard2018MNRAS}
{Bernard}, E.~J., {Schultheis}, M., {Di Matteo}, P., {et~al.} 2018, \mnras,
  477, 3507

\bibitem[{{Breddels} \& {Veljanoski}(2018)}]{2018A&A...618A..13B}
{Breddels}, M.~A. \& {Veljanoski}, J. 2018, \aap, 618, A13

\bibitem[{{Buder} {et~al.}(2021){Buder}, {Sharma}, {Kos}, {Amarsi},
  {Nordlander}, {Lind}, {Martell}, {Asplund}, {Bland-Hawthorn}, {Casey}, {de
  Silva}, {D'Orazi}, {Freeman}, {Hayden}, {Lewis}, {Lin}, {Schlesinger},
  {Simpson}, {Stello}, {Zucker}, {Zwitter}, {Beeson}, {Buck}, {Casagrande},
  {Clark}, {{\v{C}}otar}, {da Costa}, {de Grijs}, {Feuillet}, {Horner},
  {Kafle}, {Khanna}, {Kobayashi}, {Liu}, {Montet}, {Nandakumar}, {Nataf},
  {Ness}, {Spina}, {Tepper-Garc{\'\i}a}, {Ting}, {Traven},
  {Vogrin{\v{c}}i{\v{c}}}, {Wittenmyer}, {Wyse}, {{\v{Z}}erjal}, \& {GALAH
  Collaboration}}]{2021MNRAS.506..150B}
{Buder}, S., {Sharma}, S., {Kos}, J., {et~al.} 2021, \mnras, 506, 150

\bibitem[{{Bullock} \& {Johnston}(2005)}]{2005ApJ...635..931B}
{Bullock}, J.~S. \& {Johnston}, K.~V. 2005, \apj, 635, 931

\bibitem[{{Cignoni} \& {Tosi}(2010)}]{cignonitosi2010}
{Cignoni}, M. \& {Tosi}, M. 2010, Advances in Astronomy, 2010, 158568

\bibitem[{{Dodd} {et~al.}(2022){Dodd}, {Helmi}, \&
  {Koppelman}}]{2022A&A...659A..61D}
{Dodd}, E., {Helmi}, A., \& {Koppelman}, H.~H. 2022, \aap, 659, A61

\bibitem[{{Everall} \& {Boubert}(2022)}]{2022MNRAS.509.6205E}
{Everall}, A. \& {Boubert}, D. 2022, \mnras, 509, 6205

\bibitem[{{Gaia Collaboration} {et~al.}(2018){Gaia Collaboration}, {Babusiaux},
  {van Leeuwen}, {Barstow}, {Jordi}, {Vallenari}, {Bossini}, {Bressan},
  {Cantat-Gaudin}, {van Leeuwen}, {Brown}, {Prusti}, {de Bruijne},
  {Bailer-Jones}, {Biermann}, {Evans}, {Eyer}, {Jansen}, {Klioner}, {Lammers},
  {Lindegren}, {Luri}, {Mignard}, {Panem}, {Pourbaix}, {Randich}, {Sartoretti},
  {Siddiqui}, {Soubiran}, {Walton}, {Arenou}, {Bastian}, {Cropper}, {Drimmel},
  {Katz}, {Lattanzi}, {Bakker}, {Cacciari}, {Casta{\~n}eda}, {Chaoul}, {Cheek},
  {De Angeli}, {Fabricius}, {Guerra}, {Holl}, {Masana}, {Messineo}, {Mowlavi},
  {Nienartowicz}, {Panuzzo}, {Portell}, {Riello}, {Seabroke}, {Tanga},
  {Th{\'e}venin}, {Gracia-Abril}, {Comoretto}, {Garcia-Reinaldos}, {Teyssier},
  {Altmann}, {Andrae}, {Audard}, {Bellas-Velidis}, {Benson}, {Berthier},
  {Blomme}, {Burgess}, {Busso}, {Carry}, {Cellino}, {Clementini}, {Clotet},
  {Creevey}, {Davidson}, {De Ridder}, {Delchambre}, {Dell'Oro}, {Ducourant},
  {Fern{\'a}ndez-Hern{\'a}ndez}, {Fouesneau}, {Fr{\'e}mat}, {Galluccio},
  {Garc{\'\i}a-Torres}, {Gonz{\'a}lez-N{\'u}{\~n}ez}, {Gonz{\'a}lez-Vidal},
  {Gosset}, {Guy}, {Halbwachs}, {Hambly}, {Harrison}, {Hern{\'a}ndez},
  {Hestroffer}, {Hodgkin}, {Hutton}, {Jasniewicz}, {Jean-Antoine-Piccolo},
  {Jordan}, {Korn}, {Krone-Martins}, {Lanzafame}, {Lebzelter}, {L{\"o}ffler},
  {Manteiga}, {Marrese}, {Mart{\'\i}n-Fleitas}, {Moitinho}, {Mora}, {Muinonen},
  {Osinde}, {Pancino}, {Pauwels}, {Petit}, {Recio-Blanco}, {Richards},
  {Rimoldini}, {Robin}, {Sarro}, {Siopis}, {Smith}, {Sozzetti}, {S{\"u}veges},
  {Torra}, {van Reeven}, {Abbas}, {Abreu Aramburu}, {Accart}, {Aerts},
  {Altavilla}, {{\'A}lvarez}, {Alvarez}, {Alves}, {Anderson}, {Andrei},
  {Anglada Varela}, {Antiche}, {Antoja}, {Arcay}, {Astraatmadja}, {Bach},
  {Baker}, {Balaguer-N{\'u}{\~n}ez}, {Balm}, {Barache}, {Barata}, {Barbato},
  {Barblan}, {Barklem}, {Barrado}, {Barros}, {Bartholom{\'e} Mu{\~n}oz},
  {Bassilana}, {Becciani}, {Bellazzini}, {Berihuete}, {Bertone}, {Bianchi},
  {Bienaym{\'e}}, {Blanco-Cuaresma}, {Boch}, {Boeche}, {Bombrun}, {Borrachero},
  {Bouquillon}, {Bourda}, {Bragaglia}, {Bramante}, {Breddels}, {Brouillet},
  {Br{\"u}semeister}, {Brugaletta}, {Bucciarelli}, {Burlacu}, {Busonero},
  {Butkevich}, {Buzzi}, {Caffau}, {Cancelliere}, {Cannizzaro}, {Carballo},
  {Carlucci}, {Carrasco}, {Casamiquela}, {Castellani}, {Castro-Ginard},
  {Charlot}, {Chemin}, {Chiavassa}, {Cocozza}, {Costigan}, {Cowell}, {Crifo},
  {Crosta}, {Crowley}, {Cuypers}, {Dafonte}, {Damerdji}, {Dapergolas}, {David},
  {David}, {de Laverny}, {De Luise}, {De March}, {de Martino}, {de Souza}, {de
  Torres}, {Debosscher}, {del Pozo}, {Delbo}, {Delgado}, {Delgado}, {Diakite},
  {Diener}, {Distefano}, {Dolding}, {Drazinos}, {Dur{\'a}n}, {Edvardsson},
  {Enke}, {Eriksson}, {Esquej}, {Eynard Bontemps}, {Fabre}, {Fabrizio},
  {Faigler}, {Falc{\~a}o}, {Farr{\`a}s Casas}, {Federici}, {Fedorets},
  {Fernique}, {Figueras}, {Filippi}, {Findeisen}, {Fonti}, {Fraile}, {Fraser},
  {Fr{\'e}zouls}, {Gai}, {Galleti}, {Garabato}, {Garc{\'\i}a-Sedano},
  {Garofalo}, {Garralda}, {Gavel}, {Gavras}, {Gerssen}, {Geyer}, {Giacobbe},
  {Gilmore}, {Girona}, {Giuffrida}, {Glass}, {Gomes}, {Granvik}, {Gueguen},
  {Guerrier}, {Guiraud}, {Guti{\'e}}, {Haigron}, {Hatzidimitriou}, {Hauser},
  {Haywood}, {Heiter}, {Helmi}, {Heu}, {Hilger}, {Hobbs}, {Hofmann}, {Holland},
  {Huckle}, {Hypki}, {Icardi}, {Jan{\ss}en}, {Jevardat de Fombelle}, {Jonker},
  {Juh{\'a}sz}, {Julbe}, {Karampelas}, {Kewley}, {Klar}, {Kochoska}, {Kohley},
  {Kolenberg}, {Kontizas}, {Kontizas}, {Koposov}, {Kordopatis},
  {Kostrzewa-Rutkowska}, {Koubsky}, {Lambert}, {Lanza}, {Lasne}, {Lavigne}, {Le
  Fustec}, {Le Poncin-Lafitte}, {Lebreton}, {Leccia}, {Leclerc},
  {Lecoeur-Taibi}, {Lenhardt}, {Leroux}, {Liao}, {Licata}, {Lindstr{\o}m},
  {Lister}, {Livanou}, {Lobel}, {L{\'o}pez}, {Managau}, {Mann}, {Mantelet},
  {Marchal}, {Marchant}, {Marconi}, {Marinoni}, {Marschalk{\'o}}, {Marshall},
  {Martino}, {Marton}, {Mary}, {Massari}, {Matijevi{\v{c}}}, {Mazeh},
  {McMillan}, {Messina}, {Michalik}, {Millar}, {Molina}, {Molinaro},
  {Moln{\'a}r}, {Montegriffo}, {Mor}, {Morbidelli}, {Morel}, {Morris},
  {Mulone}, {Muraveva}, {Musella}, {Nelemans}, {Nicastro}, {Noval},
  {O'Mullane}, {Ord{\'e}novic}, {Ord{\'o}{\~n}ez-Blanco}, {Osborne}, {Pagani},
  {Pagano}, {Pailler}, {Palacin}, {Palaversa}, {Panahi}, {Pawlak},
  {Piersimoni}, {Pineau}, {Plachy}, {Plum}, {Poggio}, {Poujoulet},
  {Pr{\v{s}}a}, {Pulone}, {Racero}, {Ragaini}, {Rambaux}, {Ramos-Lerate},
  {Regibo}, {Reyl{\'e}}, {Riclet}, {Ripepi}, {Riva}, {Rivard}, {Rixon},
  {Roegiers}, {Roelens}, {Romero-G{\'o}mez}, {Rowell}, {Royer}, {Ruiz-Dern},
  {Sadowski}, {Sagrist{\`a} Sell{\'e}s}, {Sahlmann}, {Salgado}, {Salguero},
  {Sanna}, {Santana-Ros}, {Sarasso}, {Savietto}, {Schultheis}, {Sciacca},
  {Segol}, {Segovia}, {S{\'e}gransan}, {Shih}, {Siltala}, {Silva}, {Smart},
  {Smith}, {Solano}, {Solitro}, {Sordo}, {Soria Nieto}, {Souchay}, {Spagna},
  {Spoto}, {Stampa}, {Steele}, {Steidelm{\"u}ller}, {Stephenson}, {Stoev},
  {Suess}, {Surdej}, {Szabados}, {Szegedi-Elek}, {Tapiador}, {Taris}, {Tauran},
  {Taylor}, {Teixeira}, {Terrett}, {Teyssandier}, {Thuillot}, {Titarenko},
  {Torra Clotet}, {Turon}, {Ulla}, {Utrilla}, {Uzzi}, {Vaillant}, {Valentini},
  {Valette}, {van Elteren}, {Van Hemelryck}, {Vaschetto}, {Vecchiato},
  {Veljanoski}, {Viala}, {Vicente}, {Vogt}, {von Essen}, {Voss}, {Votruba},
  {Voutsinas}, {Walmsley}, {Weiler}, {Wertz}, {Wevers}, {Wyrzykowski},
  {Yoldas}, {{\v{Z}}erjal}, {Ziaeepour}, {Zorec}, {Zschocke}, {Zucker},
  {Zurbach}, \& {Zwitter}}]{2018A&A...616A..10G}
{Gaia Collaboration}, {Babusiaux}, C., {van Leeuwen}, F., {et~al.} 2018, \aap,
  616, A10

\bibitem[{{Gaia Collaboration} {et~al.}(2021){Gaia Collaboration}, {Brown},
  {Vallenari}, {Prusti}, {de Bruijne}, {Babusiaux}, {Biermann}, {Creevey},
  {Evans}, {Eyer}, {Hutton}, {Jansen}, {Jordi}, {Klioner}, {Lammers},
  {Lindegren}, {Luri}, {Mignard}, {Panem}, {Pourbaix}, {Randich}, {Sartoretti},
  {Soubiran}, {Walton}, {Arenou}, {Bailer-Jones}, {Bastian}, {Cropper},
  {Drimmel}, {Katz}, {Lattanzi}, {van Leeuwen}, {Bakker}, {Cacciari},
  {Casta{\~n}eda}, {De Angeli}, {Ducourant}, {Fabricius}, {Fouesneau},
  {Fr{\'e}mat}, {Guerra}, {Guerrier}, {Guiraud}, {Jean-Antoine Piccolo},
  {Masana}, {Messineo}, {Mowlavi}, {Nicolas}, {Nienartowicz}, {Pailler},
  {Panuzzo}, {Riclet}, {Roux}, {Seabroke}, {Sordo}, {Tanga}, {Th{\'e}venin},
  {Gracia-Abril}, {Portell}, {Teyssier}, {Altmann}, {Andrae}, {Bellas-Velidis},
  {Benson}, {Berthier}, {Blomme}, {Brugaletta}, {Burgess}, {Busso}, {Carry},
  {Cellino}, {Cheek}, {Clementini}, {Damerdji}, {Davidson}, {Delchambre},
  {Dell'Oro}, {Fern{\'a}ndez-Hern{\'a}ndez}, {Galluccio}, {Garc{\'\i}a-Lario},
  {Garcia-Reinaldos}, {Gonz{\'a}lez-N{\'u}{\~n}ez}, {Gosset}, {Haigron},
  {Halbwachs}, {Hambly}, {Harrison}, {Hatzidimitriou}, {Heiter},
  {Hern{\'a}ndez}, {Hestroffer}, {Hodgkin}, {Holl}, {Jan{\ss}en}, {Jevardat de
  Fombelle}, {Jordan}, {Krone-Martins}, {Lanzafame}, {L{\"o}ffler}, {Lorca},
  {Manteiga}, {Marchal}, {Marrese}, {Moitinho}, {Mora}, {Muinonen}, {Osborne},
  {Pancino}, {Pauwels}, {Petit}, {Recio-Blanco}, {Richards}, {Riello},
  {Rimoldini}, {Robin}, {Roegiers}, {Rybizki}, {Sarro}, {Siopis}, {Smith},
  {Sozzetti}, {Ulla}, {Utrilla}, {van Leeuwen}, {van Reeven}, {Abbas}, {Abreu
  Aramburu}, {Accart}, {Aerts}, {Aguado}, {Ajaj}, {Altavilla}, {{\'A}lvarez},
  {{\'A}lvarez Cid-Fuentes}, {Alves}, {Anderson}, {Anglada Varela}, {Antoja},
  {Audard}, {Baines}, {Baker}, {Balaguer-N{\'u}{\~n}ez}, {Balbinot}, {Balog},
  {Barache}, {Barbato}, {Barros}, {Barstow}, {Bartolom{\'e}}, {Bassilana},
  {Bauchet}, {Baudesson-Stella}, {Becciani}, {Bellazzini}, {Bernet}, {Bertone},
  {Bianchi}, {Blanco-Cuaresma}, {Boch}, {Bombrun}, {Bossini}, {Bouquillon},
  {Bragaglia}, {Bramante}, {Breedt}, {Bressan}, {Brouillet}, {Bucciarelli},
  {Burlacu}, {Busonero}, {Butkevich}, {Buzzi}, {Caffau}, {Cancelliere},
  {C{\'a}novas}, {Cantat-Gaudin}, {Carballo}, {Carlucci}, {Carnerero},
  {Carrasco}, {Casamiquela}, {Castellani}, {Castro-Ginard}, {Castro Sampol},
  {Chaoul}, {Charlot}, {Chemin}, {Chiavassa}, {Cioni}, {Comoretto}, {Cooper},
  {Cornez}, {Cowell}, {Crifo}, {Crosta}, {Crowley}, {Dafonte}, {Dapergolas},
  {David}, {David}, {de Laverny}, {De Luise}, {De March}, {De Ridder}, {de
  Souza}, {de Teodoro}, {de Torres}, {del Peloso}, {del Pozo}, {Delbo},
  {Delgado}, {Delgado}, {Delisle}, {Di Matteo}, {Diakite}, {Diener},
  {Distefano}, {Dolding}, {Eappachen}, {Edvardsson}, {Enke}, {Esquej}, {Fabre},
  {Fabrizio}, {Faigler}, {Fedorets}, {Fernique}, {Fienga}, {Figueras},
  {Fouron}, {Fragkoudi}, {Fraile}, {Franke}, {Gai}, {Garabato},
  {Garcia-Gutierrez}, {Garc{\'\i}a-Torres}, {Garofalo}, {Gavras}, {Gerlach},
  {Geyer}, {Giacobbe}, {Gilmore}, {Girona}, {Giuffrida}, {Gomel}, {Gomez},
  {Gonzalez-Santamaria}, {Gonz{\'a}lez-Vidal}, {Granvik},
  {Guti{\'e}rrez-S{\'a}nchez}, {Guy}, {Hauser}, {Haywood}, {Helmi}, {Hidalgo},
  {Hilger}, {H{\l}adczuk}, {Hobbs}, {Holland}, {Huckle}, {Jasniewicz},
  {Jonker}, {Juaristi Campillo}, {Julbe}, {Karbevska}, {Kervella}, {Khanna},
  {Kochoska}, {Kontizas}, {Kordopatis}, {Korn}, {Kostrzewa-Rutkowska},
  {Kruszy{\'n}ska}, {Lambert}, {Lanza}, {Lasne}, {Le Campion}, {Le Fustec},
  {Lebreton}, {Lebzelter}, {Leccia}, {Leclerc}, {Lecoeur-Taibi}, {Liao},
  {Licata}, {Lindstr{\o}m}, {Lister}, {Livanou}, {Lobel}, {Madrero Pardo},
  {Managau}, {Mann}, {Marchant}, {Marconi}, {Marcos Santos}, {Marinoni},
  {Marocco}, {Marshall}, {Martin Polo}, {Mart{\'\i}n-Fleitas}, {Masip},
  {Massari}, {Mastrobuono-Battisti}, {Mazeh}, {McMillan}, {Messina},
  {Michalik}, {Millar}, {Mints}, {Molina}, {Molinaro}, {Moln{\'a}r},
  {Montegriffo}, {Mor}, {Morbidelli}, {Morel}, {Morris}, {Mulone}, {Munoz},
  {Muraveva}, {Murphy}, {Musella}, {Noval}, {Ord{\'e}novic}, {Orr{\`u}},
  {Osinde}, {Pagani}, {Pagano}, {Palaversa}, {Palicio}, {Panahi}, {Pawlak},
  {Pe{\~n}alosa Esteller}, {Penttil{\"a}}, {Piersimoni}, {Pineau}, {Plachy},
  {Plum}, {Poggio}, {Poretti}, {Poujoulet}, {Pr{\v{s}}a}, {Pulone}, {Racero},
  {Ragaini}, {Rainer}, {Raiteri}, {Rambaux}, {Ramos}, {Ramos-Lerate}, {Re
  Fiorentin}, {Regibo}, {Reyl{\'e}}, {Ripepi}, {Riva}, {Rixon}, {Robichon},
  {Robin}, {Roelens}, {Rohrbasser}, {Romero-G{\'o}mez}, {Rowell}, {Royer},
  {Rybicki}, {Sadowski}, {Sagrist{\`a} Sell{\'e}s}, {Sahlmann}, {Salgado},
  {Salguero}, {Samaras}, {Sanchez Gimenez}, {Sanna}, {Santove{\~n}a},
  {Sarasso}, {Schultheis}, {Sciacca}, {Segol}, {Segovia}, {S{\'e}gransan},
  {Semeux}, {Shahaf}, {Siddiqui}, {Siebert}, {Siltala}, {Slezak}, {Smart},
  {Solano}, {Solitro}, {Souami}, {Souchay}, {Spagna}, {Spoto}, {Steele},
  {Steidelm{\"u}ller}, {Stephenson}, {S{\"u}veges}, {Szabados}, {Szegedi-Elek},
  {Taris}, {Tauran}, {Taylor}, {Teixeira}, {Thuillot}, {Tonello}, {Torra},
  {Torra}, {Turon}, {Unger}, {Vaillant}, {van Dillen}, {Vanel}, {Vecchiato},
  {Viala}, {Vicente}, {Voutsinas}, {Weiler}, {Wevers}, {Wyrzykowski}, {Yoldas},
  {Yvard}, {Zhao}, {Zorec}, {Zucker}, {Zurbach}, \&
  {Zwitter}}]{2021A&A...649A...1G}
{Gaia Collaboration}, {Brown}, A.~G.~A., {Vallenari}, A., {et~al.} 2021, \aap,
  649, A1

\bibitem[{{Gallart} {et~al.}(2019){Gallart}, {Bernard}, {Brook}, {Ruiz-Lara},
  {Cassisi}, {Hill}, \& {Monelli}}]{2019NatAs...3..932G}
{Gallart}, C., {Bernard}, E.~J., {Brook}, C.~B., {et~al.} 2019, Nature
  Astronomy, 3, 932

\bibitem[{{Gallart} {et~al.}(1999){Gallart}, {Freedman}, {Aparicio},
  {Bertelli}, \& {Chiosi}}]{1999AJ....118.2245G}
{Gallart}, C., {Freedman}, W.~L., {Aparicio}, A., {Bertelli}, G., \& {Chiosi},
  C. 1999, \aj, 118, 2245

\bibitem[{{Green}(2018)}]{2018JOSS....3..695M}
{Green}, G. 2018, The Journal of Open Source Software, 3, 695

\bibitem[{{Green} {et~al.}(2019){Green}, {Schlafly}, {Zucker}, {Speagle}, \&
  {Finkbeiner}}]{2019ApJ...887...93G}
{Green}, G.~M., {Schlafly}, E., {Zucker}, C., {Speagle}, J.~S., \&
  {Finkbeiner}, D. 2019, \apj, 887, 93

\bibitem[{Green \& Sibson(1978)}]{10.1093/comjnl/21.2.168}
Green, P.~J. \& Sibson, R. 1978, The Computer Journal, 21, 168

\bibitem[{{Gull} {et~al.}(2021){Gull}, {Frebel}, {Hinojosa}, {Roederer}, {Ji},
  \& {Brauer}}]{2021ApJ...912...52G}
{Gull}, M., {Frebel}, A., {Hinojosa}, K., {et~al.} 2021, \apj, 912, 52

\bibitem[{{Helmi}(2020)}]{2020ARA&A..58..205H}
{Helmi}, A. 2020, \araa, 58, 205

\bibitem[{{Helmi} {et~al.}(2018){Helmi}, {Babusiaux}, {Koppelman}, {Massari},
  {Veljanoski}, \& {Brown}}]{2018Natur.563...85H}
{Helmi}, A., {Babusiaux}, C., {Koppelman}, H.~H., {et~al.} 2018, \nat, 563, 85

\bibitem[{{Helmi} \& {de Zeeuw}(2000)}]{2000MNRAS.319..657H}
{Helmi}, A. \& {de Zeeuw}, P.~T. 2000, \mnras, 319, 657

\bibitem[{{Helmi} {et~al.}(1999){Helmi}, {White}, {de Zeeuw}, \&
  {Zhao}}]{1999Natur.402...53H}
{Helmi}, A., {White}, S. D.~M., {de Zeeuw}, P.~T., \& {Zhao}, H. 1999, \nat,
  402, 53

\bibitem[{{Hidalgo} {et~al.}(2018){Hidalgo}, {Pietrinferni}, {Cassisi},
  {Salaris}, {Mucciarelli}, {Savino}, {Aparicio}, {Silva Aguirre}, \&
  {Verma}}]{2018ApJ...856..125H}
{Hidalgo}, S.~L., {Pietrinferni}, A., {Cassisi}, S., {et~al.} 2018, \apj, 856,
  125

\bibitem[{{Hirschmann} {et~al.}(2012){Hirschmann}, {Naab}, {Somerville},
  {Burkert}, \& {Oser}}]{2012MNRAS.419.3200H}
{Hirschmann}, M., {Naab}, T., {Somerville}, R.~S., {Burkert}, A., \& {Oser}, L.
  2012, \mnras, 419, 3200

\bibitem[{{Horta} {et~al.}(2021){Horta}, {Schiavon}, {Mackereth}, {Pfeffer},
  {Mason}, {Kisku}, {Fragkoudi}, {Allende Prieto}, {Cunha}, {Hasselquist},
  {Holtzman}, {Majewski}, {Nataf}, {O'Connell}, {Schultheis}, \&
  {Smith}}]{2021MNRAS.500.1385H}
{Horta}, D., {Schiavon}, R.~P., {Mackereth}, J.~T., {et~al.} 2021, \mnras, 500,
  1385

\bibitem[{{Horta} {et~al.}(2022){Horta}, {Schiavon}, {Mackereth}, {Weinberg},
  {Hasselquist}, {Feuillet}, {O'Connell}, {Anguiano}, {Allende-Prieto},
  {Beaton}, {Bizyaev}, {Cunha}, {Geisler}, {Garc{\'\i}a-Hern{\'a}ndez},
  {Holtzman}, {J{\"o}nsson}, {Lane}, {Majewski}, {M{\'e}sz{\'a}ros}, {Minniti},
  {Nitschelm}, {Shetrone}, {Smith}, \& {Zasowski}}]{2022arXiv220404233H}
{Horta}, D., {Schiavon}, R.~P., {Mackereth}, J.~T., {et~al.} 2022, arXiv
  e-prints, arXiv:2204.04233

\bibitem[{Hunter(2007)}]{hunter2007}
Hunter, J.~D. 2007, Computing In Science \& Engineering, 9, 90

\bibitem[{{Ibata} {et~al.}(1994){Ibata}, {Gilmore}, \&
  {Irwin}}]{1994Natur.370..194I}
{Ibata}, R.~A., {Gilmore}, G., \& {Irwin}, M.~J. 1994, \nat, 370, 194

\bibitem[{{Kepley} {et~al.}(2007){Kepley}, {Morrison}, {Helmi}, {Kinman}, {Van
  Duyne}, {Martin}, {Harding}, {Norris}, \& {Freeman}}]{2007AJ....134.1579K}
{Kepley}, A.~A., {Morrison}, H.~L., {Helmi}, A., {et~al.} 2007, \aj, 134, 1579

\bibitem[{{Koppelman} \& {Helmi}(2021)}]{2021A&A...649A.136K}
{Koppelman}, H.~H. \& {Helmi}, A. 2021, \aap, 649, A136

\bibitem[{{Koppelman} {et~al.}(2019{\natexlab{a}}){Koppelman}, {Helmi},
  {Massari}, {Price-Whelan}, \& {Starkenburg}}]{2019A&A...631L...9K}
{Koppelman}, H.~H., {Helmi}, A., {Massari}, D., {Price-Whelan}, A.~M., \&
  {Starkenburg}, T.~K. 2019{\natexlab{a}}, \aap, 631, L9

\bibitem[{{Koppelman} {et~al.}(2019{\natexlab{b}}){Koppelman}, {Helmi},
  {Massari}, {Roelenga}, \& {Bastian}}]{2019A&A...625A...5K}
{Koppelman}, H.~H., {Helmi}, A., {Massari}, D., {Roelenga}, S., \& {Bastian},
  U. 2019{\natexlab{b}}, \aap, 625, A5

\bibitem[{{Kormendy} \& {Kennicutt}(2004)}]{2004ARA&A..42..603K}
{Kormendy}, J. \& {Kennicutt}, Robert~C., J. 2004, \araa, 42, 603

\bibitem[{{Kroupa}(2001)}]{kroupa2001}
{Kroupa}, P. 2001, \mnras, 322, 231

\bibitem[{{Limberg} {et~al.}(2021){Limberg}, {Santucci}, {Rossi}, {Queiroz},
  {Chiappini}, {Souza}, {Perottoni}, {P{\'e}rez-Villegas}, \&
  {Barbosa}}]{2021ApJ...913L..28L}
{Limberg}, G., {Santucci}, R.~M., {Rossi}, S., {et~al.} 2021, \apjl, 913, L28

\bibitem[{{Lindegren} {et~al.}(2021){Lindegren}, {Bastian}, {Biermann},
  {Bombrun}, {de Torres}, {Gerlach}, {Geyer}, {Hern{\'a}ndez}, {Hilger},
  {Hobbs}, {Klioner}, {Lammers}, {McMillan}, {Ramos-Lerate},
  {Steidelm{\"u}ller}, {Stephenson}, \& {van Leeuwen}}]{2021A&A...649A...4L}
{Lindegren}, L., {Bastian}, U., {Biermann}, M., {et~al.} 2021, \aap, 649, A4

\bibitem[{{Liu} {et~al.}(2019){Liu}, {Fu}, {Zong}, {Shi}, {Luo}, {Zhang},
  {Cui}, {Hou}, {Pan}, {Shan}, {Chen}, {Bai}, {Chen}, {Du}, {Hou}, {Liu},
  {Tian}, {Wang}, {Wang}, {Wu}, {Wu}, {Yan}, \& {Zuo}}]{2019RAA....19...75L}
{Liu}, N., {Fu}, J.-N., {Zong}, W., {et~al.} 2019, Research in Astronomy and
  Astrophysics, 19, 075

\bibitem[{{L{\"o}vdal} {et~al.}(2022){L{\"o}vdal}, {Ruiz-Lara}, {Koppelman},
  {Matsuno}, {Dodd}, \& {Helmi}}]{2022A&A...665A..57L}
{L{\"o}vdal}, S.~S., {Ruiz-Lara}, T., {Koppelman}, H.~H., {et~al.} 2022, \aap,
  665, A57

\bibitem[{{Matsuno} {et~al.}(2022){Matsuno}, {Dodd}, {Koppelman}, {Helmi},
  {Ishigaki}, {Aoki}, {Zhao}, {Yuan}, \& {Hattori}}]{2022arXiv220311808M}
{Matsuno}, T., {Dodd}, E., {Koppelman}, H.~H., {et~al.} 2022, arXiv e-prints,
  arXiv:2203.11808

\bibitem[{{Matsuno} {et~al.}(2021){Matsuno}, {Koppelman}, {Helmi}, {Aoki},
  {Ishigaki}, {Suda}, {Yuan}, \& {Hattori}}]{2021arXiv211115423M}
{Matsuno}, T., {Koppelman}, H.~H., {Helmi}, A., {et~al.} 2021, arXiv e-prints,
  arXiv:2111.15423

\bibitem[{{McMillan}(2017)}]{2017MNRAS.465...76M}
{McMillan}, P.~J. 2017, \mnras, 465, 76

\bibitem[{{Monelli} {et~al.}(2010){Monelli}, {Hidalgo}, {Stetson}, {Aparicio},
  {Gallart}, {Dolphin}, {Cole}, {Weisz}, {Skillman}, {Bernard}, {Mayer},
  {Navarro}, {Cassisi}, {Drozdovsky}, \& {Tolstoy}}]{monelli2010}
{Monelli}, M., {Hidalgo}, S.~L., {Stetson}, P.~B., {et~al.} 2010, \apj, 720,
  1225

\bibitem[{{Mor} {et~al.}(2019){Mor}, {Robin}, {Figueras}, {Roca-F{\`a}brega},
  \& {Luri}}]{2019A&A...624L...1M}
{Mor}, R., {Robin}, A.~C., {Figueras}, F., {Roca-F{\`a}brega}, S., \& {Luri},
  X. 2019, \aap, 624, L1

\bibitem[{{Myeong} {et~al.}(2018){Myeong}, {Evans}, {Belokurov}, {Sanders}, \&
  {Koposov}}]{2018MNRAS.478.5449M}
{Myeong}, G.~C., {Evans}, N.~W., {Belokurov}, V., {Sanders}, J.~L., \&
  {Koposov}, S.~E. 2018, \mnras, 478, 5449

\bibitem[{{Naidu} {et~al.}(2020){Naidu}, {Conroy}, {Bonaca}, {Johnson}, {Ting},
  {Caldwell}, {Zaritsky}, \& {Cargile}}]{2020ApJ...901...48N}
{Naidu}, R.~P., {Conroy}, C., {Bonaca}, A., {et~al.} 2020, \apj, 901, 48

\bibitem[{{Naidu} {et~al.}(2022){Naidu}, {Conroy}, {Bonaca}, {Zaritsky},
  {Ting}, {Caldwell}, {Cargile}, {Speagle}, {Chandra}, {Johnson}, {Woody}, \&
  {Han}}]{2022arXiv220409057N}
{Naidu}, R.~P., {Conroy}, C., {Bonaca}, A., {et~al.} 2022, arXiv e-prints,
  arXiv:2204.09057

\bibitem[{{Nissen} {et~al.}(2021){Nissen}, {Silva-Cabrera}, \&
  {Schuster}}]{2021A&A...651A..57N}
{Nissen}, P.~E., {Silva-Cabrera}, J.~S., \& {Schuster}, W.~J. 2021, \aap, 651,
  A57

\bibitem[{{Pearson} {et~al.}(2019){Pearson}, {Wang}, {Alpaslan}, {Baldry},
  {Bilicki}, {Brown}, {Grootes}, {Holwerda}, {Kitching}, {Kruk}, \& {van der
  Tak}}]{2019A&A...631A..51P}
{Pearson}, W.~J., {Wang}, L., {Alpaslan}, M., {et~al.} 2019, \aap, 631, A51

\bibitem[{{Pietrinferni} {et~al.}(2004){Pietrinferni}, {Cassisi}, {Salaris}, \&
  {Castelli}}]{2004ApJ...612..168P}
{Pietrinferni}, A., {Cassisi}, S., {Salaris}, M., \& {Castelli}, F. 2004, \apj,
  612, 168

\bibitem[{{Pietrinferni} {et~al.}(2021){Pietrinferni}, {Hidalgo}, {Cassisi},
  {Salaris}, {Savino}, {Mucciarelli}, {Verma}, {Silva Aguirre}, {Aparicio}, \&
  {Ferguson}}]{2021ApJ...908..102P}
{Pietrinferni}, A., {Hidalgo}, S., {Cassisi}, S., {et~al.} 2021, \apj, 908, 102

\bibitem[{{Roederer} {et~al.}(2010){Roederer}, {Sneden}, {Thompson}, {Preston},
  \& {Shectman}}]{2010ApJ...711..573R}
{Roederer}, I.~U., {Sneden}, C., {Thompson}, I.~B., {Preston}, G.~W., \&
  {Shectman}, S.~A. 2010, \apj, 711, 573

\bibitem[{{Ruiz-Lara} {et~al.}(2020){Ruiz-Lara}, {Gallart}, {Bernard}, \&
  {Cassisi}}]{2020NatAs...4..965R}
{Ruiz-Lara}, T., {Gallart}, C., {Bernard}, E.~J., \& {Cassisi}, S. 2020, Nature
  Astronomy, 4, 965

\bibitem[{{Ruiz-Lara} {et~al.}(2021){Ruiz-Lara}, {Gallart}, {Monelli}, {Fritz},
  {Battaglia}, {Cassisi}, {Aznar}, {Russo Cabrera},
  {Rodr{\'\i}guez-Mart{\'\i}n}, \&
  {Salazar-Gonz{\'a}lez}}]{2021MNRAS.501.3962R}
{Ruiz-Lara}, T., {Gallart}, C., {Monelli}, M., {et~al.} 2021, \mnras, 501, 3962

\bibitem[{{Ruiz-Lara} {et~al.}(2022){Ruiz-Lara}, {Matsuno}, {L{\"o}vdal},
  {Helmi}, {Dodd}, \& {Koppelman}}]{2022arXiv220102405R}
{Ruiz-Lara}, T., {Matsuno}, T., {L{\"o}vdal}, S.~S., {et~al.} 2022, arXiv
  e-prints, arXiv:2201.02405

\bibitem[{{Sancisi} {et~al.}(2008){Sancisi}, {Fraternali}, {Oosterloo}, \& {van
  der Hulst}}]{2008A&ARv..15..189S}
{Sancisi}, R., {Fraternali}, F., {Oosterloo}, T., \& {van der Hulst}, T. 2008,
  \aapr, 15, 189

\bibitem[{{Searle} \& {Zinn}(1978)}]{1978ApJ...225..357S}
{Searle}, L. \& {Zinn}, R. 1978, \apj, 225, 357

\bibitem[{{Steinmetz} {et~al.}(2020{\natexlab{a}}){Steinmetz}, {Guiglion},
  {McMillan}, {Matijevi{\v{c}}}, {Enke}, {Kordopatis}, {Zwitter}, {Valentini},
  {Chiappini}, {Casagrande}, {Wojno}, {Anguiano}, {Bienaym{\'e}}, {Bijaoui},
  {Binney}, {Burton}, {Cass}, {de Laverny}, {Fiegert}, {Freeman}, {Fulbright},
  {Gibson}, {Gilmore}, {Grebel}, {Helmi}, {Kunder}, {Munari}, {Navarro},
  {Parker}, {Ruchti}, {Recio-Blanco}, {Reid}, {Seabroke}, {Siviero}, {Siebert},
  {Stupar}, {Watson}, {Williams}, {Wyse}, {Anders}, {Antoja}, {Birko},
  {Bland-Hawthorn}, {Bossini}, {Garc{\'\i}a}, {Carrillo}, {Chaplin},
  {Elsworth}, {Famaey}, {Gerhard}, {Jofre}, {Just}, {Mathur}, {Miglio},
  {Minchev}, {Monari}, {Mosser}, {Ritter}, {Rodrigues}, {Scholz}, {Sharma},
  {Sysoliatina}, \& {RAVE Collaboration}}]{2020AJ....160...83S}
{Steinmetz}, M., {Guiglion}, G., {McMillan}, P.~J., {et~al.}
  2020{\natexlab{a}}, \aj, 160, 83

\bibitem[{{Steinmetz} {et~al.}(2020{\natexlab{b}}){Steinmetz},
  {Matijevi{\v{c}}}, {Enke}, {Zwitter}, {Guiglion}, {McMillan}, {Kordopatis},
  {Valentini}, {Chiappini}, {Casagrande}, {Wojno}, {Anguiano}, {Bienaym{\'e}},
  {Bijaoui}, {Binney}, {Burton}, {Cass}, {de Laverny}, {Fiegert}, {Freeman},
  {Fulbright}, {Gibson}, {Gilmore}, {Grebel}, {Helmi}, {Kunder}, {Munari},
  {Navarro}, {Parker}, {Ruchti}, {Recio-Blanco}, {Reid}, {Seabroke}, {Siviero},
  {Siebert}, {Stupar}, {Watson}, {Williams}, {Wyse}, {Anders}, {Antoja},
  {Birko}, {Bland-Hawthorn}, {Bossini}, {Garc{\'\i}a}, {Carrillo}, {Chaplin},
  {Elsworth}, {Famaey}, {Gerhard}, {Jofre}, {Just}, {Mathur}, {Miglio},
  {Minchev}, {Monari}, {Mosser}, {Ritter}, {Rodrigues}, {Scholz}, {Sharma},
  {Sysoliatina}, \& {RAVE Collaboration}}]{2020AJ....160...82S}
{Steinmetz}, M., {Matijevi{\v{c}}}, G., {Enke}, H., {et~al.}
  2020{\natexlab{b}}, \aj, 160, 82

\bibitem[{{Tolstoy} {et~al.}(2009){Tolstoy}, {Hill}, \& {Tosi}}]{tolstoy2009}
{Tolstoy}, E., {Hill}, V., \& {Tosi}, M. 2009, \araa, 47, 371

\bibitem[{{Tsantaki} {et~al.}(2022){Tsantaki}, {Pancino}, {Marrese},
  {Marinoni}, {Rainer}, {Sanna}, {Turchi}, {Randich}, {Gallart}, {Battaglia},
  \& {Masseron}}]{2022A&A...659A..95T}
{Tsantaki}, M., {Pancino}, E., {Marrese}, P., {et~al.} 2022, \aap, 659, A95

\bibitem[{{Vincenzo} {et~al.}(2019){Vincenzo}, {Spitoni}, {Calura},
  {Matteucci}, {Silva Aguirre}, {Miglio}, \& {Cescutti}}]{2019MNRAS.487L..47V}
{Vincenzo}, F., {Spitoni}, E., {Calura}, F., {et~al.} 2019, \mnras, 487, L47

\bibitem[{{Wang} {et~al.}(2020){Wang}, {Fu}, {Zong}, {Smith}, {De Cat}, {Shi},
  {Luo}, {Zhang}, {Frasca}, {Corbally}, {Molenda-{\.Z}akowicz}, {Catanzaro},
  {Gray}, {Wang}, \& {Pan}}]{2020ApJS..251...27W}
{Wang}, J., {Fu}, J.-N., {Zong}, W., {et~al.} 2020, \apjs, 251, 27

\bibitem[{{Zolotov} {et~al.}(2009){Zolotov}, {Willman}, {Brooks}, {Governato},
  {Brook}, {Hogg}, {Quinn}, \& {Stinson}}]{2009ApJ...702.1058Z}
{Zolotov}, A., {Willman}, B., {Brooks}, A.~M., {et~al.} 2009, \apj, 702, 1058

\end{thebibliography}
%\input{references}

%%%%%%%%%%%%%%%%%%%%%%%%%%%%%%%%%%%%%%%%%%%%%%%%%%

%%%%%%%%%%%%%%%%% APPENDICES %%%%%%%%%%%%%%%%%%%%%

\onecolumn
\appendix

\section{Assessing the robustness of the method}
\label{app:tests}

\begin{figure}
\centering
\includegraphics[width=0.95\textwidth]{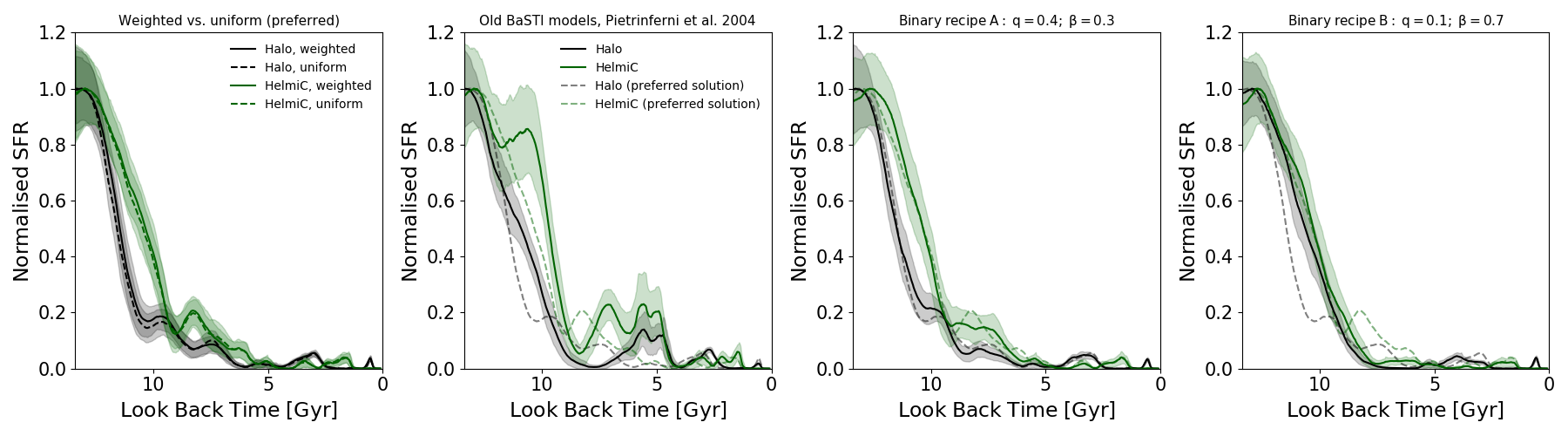}
\caption{Comparison of the SFHs of the 5D Helmi streams (green) with that of the halo star selection (black) for different configurations of {dirSFH}. From left to right, we show the effect of:  (i) the weighting scheme, (ii) different versions of BaSTI models \citep[][]{2004ApJ...612..168P}; and (iii) and (iv) different recipes for modelling unresolved binaries in the synthetic CMD. In all cases we use the weighted option except in the first panel, where we assess the effect of a uniform weighting (dashed lines).  In the three right-most panels we indicate the preferred solution discussed throughout the paper as dashed lines for comparison. All SFHs are normalised to their peak value.
\label{fig:dirSFH_tests}}
\end{figure}

Figure~\ref{fig:dirSFH_tests} displays the comparison between the SFH of the HelmiC and local retrograde halo 5D samples as obtained using different configurations of dirSFH. The left-hand panel highlights the small effect that the weighting scheme has on our solutions (for the particular case of our preferred configuration).

In all tests, we see how our main result, that the Helmi stream progenitor displays a more extended early SFH with respect to the overall halo, holds (see Fig.~\ref{fig:dirSFH_tests}). It is only minimised (differences of 0.2~Gyr in the half-mass formation time) when an extreme amount of unresolved binaries is included in the synthetic CMD ($\beta$=70$\%$). While the recovery of the HelmiC SFH  is nearly unaffected in this case, the halo solution displays lower metallicities than that found from spectroscopic surveys, which is compensated for in the dirSFH solution with younger ages (age-[Fe/H] degeneracy), and thus the differences between the SFH of the Helmi progenitor and the halo are minimised. \citet[][]{2020MNRAS.496.1922B} studied the variation in the fraction of unresolved binaries in the CMD using {\it Gaia} DR2 data, finding that, in the range of colours and magnitudes sampled in this work, the fraction of unresolved binaries is $\sim$30$\%$ (the same fraction assumed in this work). All this can be used as evidence to support our choice and the robustness of the method, and it confirms that the progenitor of the Helmi streams had a more extended period of early star formation than the MW halo.

\end{document}